\documentclass[aps,twocolumn,showpacs,floatfix,superscriptaddress]{revtex4}

\usepackage{epsfig}
\usepackage{amsmath}
\def\lsim{\raise0.3ex\hbox{$\;<$\kern-0.75em\raise-1.1ex\hbox{$\sim\;$}}}
\def\gsim{\raise0.3ex\hbox{$\;>$\kern-0.75em\raise-1.1ex\hbox{$\sim\;$}}}
\def\eps{\varepsilon}
\def\theta{\vartheta}

\newcommand{\be}{\begin{equation}}
\newcommand{\ee}{\end{equation}}
\newcommand{\ba}{\begin{eqnarray}}
\newcommand{\ea}{\end{eqnarray}}
\newcommand{\nn}{\nonumber}

\newcommand{\mc}{\mathcal}
\newcommand{\half}{{\textstyle\frac{1}{2}}}

\def\Ms{M_{\rm SUSY}}

\begin{document}

\title{Supersymmetric superheavy dark matter }

\author{V.~Berezinsky}
\affiliation{INFN, Laboratori Nazionali del Gran Sasso, I--67010
  Assergi (AQ), Italy}
\affiliation{Institute for Nuclear Research of the RAS, Moscow, Russia}

\author{M.~Kachelrie\ss}
\affiliation{Institutt for fysikk, NTNU Trondheim, N--7491 Trondheim,
  Norway}

\author{M.~Aa.~Solberg}
\affiliation{Institutt for fysikk, NTNU Trondheim, N--7491 Trondheim,
  Norway}

\date{October 16, 2008}

\begin{abstract}
We propose the lightest supersymmetric particle (LSP) as a well-suited 
candidate for superheavy dark matter (SHDM). Various production mechanisms
at the end of inflation can produce SHDM with the correct abundance, 
$\Omega_{\rm LSP} h^2 \sim 0.1$, 
if its mass is sufficiently high. In particular, gravitational production
requires that the mass $m_{\rm LSP}$ of the LSP is above 
$3\times 10^{11}\,{\rm GeV}$. Weak interactions remain perturbative despite 
the large mass hierarchy, $m_{\rm LSP}\gg m_Z$, because of the special 
decoupling properties of supersymmetry. As a result the model is predictive 
and we discuss the relevant cosmological processes for the case of a 
superheavy neutralino within this scheme.
\end{abstract}


\maketitle

\section{Introduction}
\label{introduction}

Uncovering the nature of dark matter (DM) is one of 
the most pressing problems of current research in particle physics
and cosmology. A wealth of observational data suggests that 
a viable DM candidate has to be non-baryonic and should be non-relativistic, 
at least from the  time of matter-radiation equilibrium on~\cite{DMreviews}. 
The various particles proposed as cold dark matter (CDM) candidates $X$
can be divided in two main sub-categories: 
Thermal relics were at least once during the history of the Universe in 
chemical equilibrium with the thermal plasma, while non-thermal relics
have either sufficiently small interactions or a high enough mass $m_X$ to be 
never produced efficiently by processes like, e.g., $e^-e^+\to XX$.

The present relic abundance $\Omega_X$ of a thermal relic scales 
approximately as $\Omega_X\propto 1/\sigma_{\rm ann}$ with its annihilation 
cross section $\sigma_{\rm ann}$. Moreover, unitarity of the $S$ matrix 
restricts annihilations into the $l$.th partial-wave of particles with 
relative velocity  $v_{\rm rel}$ as 
$\sigma_{\rm ann}^{(l)}\leq (2l+1)4\pi/(v_{\rm rel}M_X^2)$~\cite{GKH}.
Since for non-relativistic point-particles higher partial-waves are 
suppressed, the observed value~\cite{WMAP5} $\Omega_{\rm CDM}h^2=0.11$ 
of the DM abundance constrains the mass of any thermal relic as 
$m\lsim 100\,$TeV.
On the other hand, the requirement that the DM is cold translates for 
thermal relics into a lower mass limit of the order 10\,keV. Thus the mass 
of thermal relics  should lie in the 10\,keV -- 100\,TeV range.

Supersymmetry (SUSY) provides with the lightest neutralino one of the most 
attractive candidates for thermal DM, see Ref.~\cite{JKG} for a review.  
Low-energy SUSY models are a natural extension of the 
standard model (SM), offering a solution to the hierarchy mass problem
and a realization of electroweak symmetry breaking by radiative 
corrections~\cite{EWRSB}. The typical range of neutralino masses 
in these models extends from a few tens of GeV up to $\sim 10~$\,TeV.

Two notable {\em non-thermal DM candidates\/} are axions and superheavy
DM (SHDM) particles. Axions were proposed as solution to the strong 
CP problem, but the still viable ``axion window'' includes the possibility 
that axions are the main contribution to the DM abundance.
Depending on the inflationary scenario, axions may be non-thermally produced 
either by the misalignment mechanism or the decay of axionic topological 
defects~\cite{Sikivie}. 

We will review the status of the SHDM model in the 
next section, before we  discuss ``superheavy supersymmetry'' in Sec.~III. 
Results for the relevant cosmological processes of a superheavy neutralino 
like elastic scattering  on light fermions, self-scatterings 
and annihilations are presented in Sec.~IV. After that, we review various 
possible production mechanisms of SHDM for the special case of a superheavy 
neutralino in Sec.~V and summarize in Sec.VI.

\section{Superheavy Dark Matter}
\label{SHDM}

The first proposal of SHDM in Refs.~\cite{BKV,KR} was motivated by 
observations of ultrahigh energy cosmic rays (UHECR), which revealed 
not the expected suppression of the energy spectrum due to the interaction 
of extragalactic protons with cosmic microwave photons. Since CDM is 
gravitationally accumulated in the halo of our galaxy, the secondaries 
produced in decays or annihilations of SHDM do not suffer energy losses
and their energy spectrum is characterized by a flat spectrum up to the 
kinematical cutoff $m_X/2$. In this scenario, the mass $m_X$ of the SHDM 
particle should exceed  $10^{12}$\,GeV. Meanwhile, this original 
motivation for proposing SHDM has disappeared in the light of new UHECR 
data compatible with the expected flux suppression, for more details
see Ref.~\cite{blois}. 

Superheavy dark matter particles with the density required by cosmological
observations can be efficiently
produced at inflation by many mechanisms including thermal 
production~\cite{BKV,KR,ChKR-r}. The most
detailed description of this process within an inflationary framework
is given in Ref.~\cite{Kolb2003}. 

A variety of different production mechanisms can provide 
a non-thermal distribution of superheavy particles in the expanding
universe.  
Since the energy density of non-relativistic particles 
decreases slower than the one of radiation, their abundance increases by the 
factor $a(t_0)/a(t_\ast)$ with respect to radiation, where $a(t_0)$ and 
$a(t_\ast)$ are the scale factors of the universe today and at the epoch of 
particle generation, respectively. If particle
production happens  at the earliest relevant time, i.e.\ during
inflation, this 
factor can become extremely large, $\sim 10^{22}$. Not surprisingly, 
such a small energy fraction can be transferred to SHDM particles by
many different 
mechanisms,  as thermal production at reheating \cite{BKV,ChKR-r}, the   
non-perturbative regime of a broad parametric resonance at 
preheating~\cite{preheating,inst-preheat}, and production by topological 
defects \cite{BKV,Kolb}. 

We discuss first the generation of superheavy particles by gravitational 
interactions from vacuum at the end of inflation~\cite{grav,grav2}. Since 
this production mechanism relies only on the gravitational coupling of the 
SHDM particle, it is unavoidably present in contrast to other, more 
model-dependent generation mechanisms. Neither inflation is needed
for this production, it rather limits the gravitational production of 
the particles. Since this production is caused by the time variation of 
the Hubble parameter $H(t)$, only particles with masses 
$m_X \lsim H(t)$ can be produced. In inflationary scenarios 
$H(t) \lsim m_{\phi}$, where $m_{\phi}$ is the mass of the inflaton. It
results in the limit on the mass of the produced particles, 
$m_X \lsim 10^{13}$~GeV \cite{grav,grav2}. 

The numerical calculations of Ref.~\cite{grav2} for the present abundance 
of fermionic SHDM can be approximated as
\be \label{abund}
 \Omega_X h^2 \approx \frac{T_R}{10^{8}\,{\rm GeV}} 
           \:\left\{
           \begin{array}{ll} 
            \left( m_X/H_I \right)^2,
            & \qquad m_X\ll H_I \\
            \exp( -m_X/H_I)  \,,     
            & \qquad m_X\gg H_I 
           \end{array}
           \right.\,,
\ee
where $H_I\approx 10^{13}$\,GeV is the Hubble parameter during inflation, 
and $T_R$ is the reheating temperature. 

Other generation mechanisms can occur additionally to gravitational 
production. If the SHDM particles couples directly or through an intermediate 
particle to the inflaton field, the time-dependence of the classical
inflaton field induces particle production. SHDM particles may be also
efficiently produced at preheating~\cite{preheating}. This stage, 
predecessor of reheating, is caused by oscillations of the inflaton field 
relative to the minimum of the 
potential after inflation. Such an oscillating field can  produce
non-perturbatively in the regime of a broad parametric resonance 
intermediate bosons which then decay to SHDM particles. The mass of
the SHDM particles
can be one or even two orders of magnitude larger than the inflaton mass.

Another mechanism is the so-called instant preheating~\cite{inst-preheat}. 
It works only in specific models, where the mass of the intermediate boson 
$\chi$ is proportional to the inflaton field, $m_{\chi} = g \phi$. When the 
inflaton crosses the potential minimum $\phi=0$, $\chi$ particles are 
massless and they are efficiently produced. When $|\phi|$ increases, $m_{\chi}$
increases, too, and can reach values up to the Planck mass.

While these additional production mechanisms can increase the 
abundance of SHDM particles relative to Eq.~(\ref{abund}), entropy production
as for instance in thermal inflation can reduce their abundance.
Therefore there exists only a lower, not an upper limit for the mass of  
SHDM, arising from the condition that the SHDM particles do not reach
chemical equilibrium.

What are the particle candidates for SHDM? 

The first problem one meets is the particle {\em life-time}. 
Superheavy particles are expected to be very short-lived: Even gravitational
interactions, e.g.\ described by dimension 5 operators suppressed by the
Planck mass, result in lifetimes much shorter than the age of 
the universe $t_0$. Superheavy particles must be thus protected from
fast decays by a symmetry which is respected even by gravity. 
Such symmetries are known: They are {\em discrete gauge symmetries}. 
These symmetries can be very weakly broken, e.g. by wormhole~\cite{BKV} 
or instanton effects~\cite{KR}, to provide a sufficiently long lifetime $\tau$,
$\tau\gsim t_0$. A systematic analysis of broken discrete gauge symmetries
is given in Ref.~\cite{Ha98}. For instance, the lifetime of SHDM with 
mass $m_X \sim 10^{13} - 10^{14}$\,GeV was found to be in the
range $10^{11} - 10^{26}$\,yr in the case of the symmetry group $Z_{10}$. 

Various models that contain either absolutely stable or unstable particles 
with life-times larger than the age of the universe
have  been discussed~\cite{Ha98,crypton1,Coriano}. Most of the suggested 
SHDM candidates belong to a new sector that has no 
tree-level interactions with SM particles. By 
contrast, we study in this work the possibility of having 
a SHDM particle with SM-like couplings to the weak gauge bosons.
Since the longitudinal part of gauge bosons couples as $\propto gM_X/m_Z$ 
to a particle with mass $M_X$, weak interactions become generically strong 
for $M_X\gg m_Z$ and thus the perturbative expansion fails. Using partial-wave 
unitarity, Chanowitz, Furman and Hinchliffe~\cite{CFH} derived thereby 
an upper  limit of $M_X\sim {\rm TeV}$ for particles coupling with SM strength
to the weak gauge bosons. 
An exception to this bound are supersymmetric theories, if only mass terms
are added that break supersymmetry (SUSY) softly~\cite{soft},
and in particular the minimal supersymmetric extension of the SM 
(MSSM)~\cite{Dobado}. Therefore we are led to suggest superheavy 
supersymmetry, i.e.\ the case where all masses of supersymmetric 
particles are of order $10^{11}\,$GeV or larger, as a concrete model for
SHDM with SM weak interactions. For definiteness, we choose the LSP as the 
lightest neutralino but we note that other possibilities as a sneutrino
are also viable.

\section{Superheavy supersymmetry}
\label{SS}

Our main motivation for the introduction of superheavy supersymmetry is
the search for a particle candidate for superheavy dark matter.   
As discussed in the previous section, the particle candidates 
found so far~\cite{Ha98,crypton1,Coriano} are in  new  particle sectors, 
such as e.g.\ the hidden sectors of supergravity or string
models.  In this section we discuss a more natural and more familiar
candidate, the lightest supersymmetric particle, in the case
that all supersymmetric particles are superheavy. The longevity of this 
dark matter candidate is provided by a $Z_2$ discrete gauge symmetry 
(R-parity) and its production is guaranteed by gravitational interactions
in any standard inflationary scenario.
The scale of SUSY breaking may be determined by LHC experiments, and 
our model can be soon falsified by the discovery of low-scale SUSY
at LHC. 

Independent of the scale of symmetry breaking, supersymmetry remains an 
inevitable feature of any theory that wants to unify internal gauge symmetries
such as SU(5) or SO(10) with the symmetry group of Minkowski space-time, 
the Poincar{\'e}\ group. Promoted to a local symmetry, gravity and 
gauge interactions are on a similar footing, with the gravitino as gauge 
field of gravity. Another motivation  to consider (superheavy) SUSY is that 
it is a generic ingredient of consistent string theories. Thus various 
esthetical reasons suggest that SUSY may be realized in Nature.

Below we discuss the status of superheavy supersymmetry in comparison
with low-scale symmetry breaking.

There are three pieces 
of evidence pointing towards ``low-scale SUSY'', i.e.\ a mass scale 
$\Ms$ of the supersymmetric partners
of the SM particles below or around 1\,TeV.
First, the unification of coupling constants fails in the SM, while 
the three couplings meet in the MSSM assuming 
$\Ms\sim 1\,$TeV~\cite{coupling}. Second, the fine-tuning problem of
the SM Higgs is remedied in the MSSM, only if the mass splitting between 
the SM particles and their SUSY partners is small enough.   
Third, the SM contains no suitable DM candidate, while the lightest 
supersymmetric particle (LSP) of the MSSM, assuming that
$R$ parity is conserved, is a promising CDM candidate. Assuming further that
the LSP is a thermal relic requires again that at least part of the
SUSY mass spectrum is below or close to the TeV scale.  

These attractive properties of low-scale SUSY are overshadowed 
by several less appealing features:  Low-scale SUSY predicts generically 
excessive flavor and CP violation as well as proton decay through 
dimension-5 operators that is close to or exceeds observational bounds.
Moreover, the non-observation of SUSY particles with masses around 100\,GeV 
re-introduces a ``small'' fine-tuning problem \cite{finetuning}:
For instance, electroweak symmetry breaking requires that
\be
 \frac{m_Z^2}{2} = \frac{m_2^2-m_1^2\tan^2\beta}{\tan^2\beta-1} - \mu^2
\,,
\ee
where $m_i$ are the usual mass parameters of the Higgs potential that
depend implicitly on the soft SUSY breaking masses. For 
$m_Z^2\ll m_i^2,\mu^2$, a certain amount of cancellation between the terms 
on the RHS is required.
Similarly, an upper limit on neutralino mass  mass of 200\,GeV arises,
if one limits accidental fine tuning to the level of 1\%~\cite{Torino}.
Moreover, most of the so-called ``bulk region'' in which the lightest 
neutralino has naturally the correct DM abundance is meanwhile 
excluded~\cite{Ellis:2003cw}.

Recently, split SUSY has been proposed as a model avoiding the problems 
of low-scale SUSY while keeping gauge coupling unification~\cite{split}.
In this model, the mass spectrum of SUSY particles is separated in two parts: 
Gauginos and gluinos are kept at the TeV scale providing with the lightest
neutralino a suitable
thermal DM candidate, while all scalars additional to the SM Higgs are heavy, 
with masses possibly close to the GUT scale. Motivated by the
cosmological constant problem and the landscape picture~\cite{land} suggested 
by string theory, the naturalness principle
is given up, keeping as guiding principles only experimental observations:
The existence of DM, and the hint for GUT from gauge coupling unification.

In this work, we go one step further by abandoning also for the gaugino  
and gluino masses the weak scale. This becomes possible because
we assume that the LSP is produced non-thermally at the end of inflation.
As a result, the mass of the LSP should be generically 
above $\sim 3\times  10^{11}\,$GeV. 
The mass of the gluinos and of the SUSY scalars could be either close or, 
in a similar but not as extreme set-up as in split SUSY, much larger. 
In such a set-up, alternative approaches as modular~\cite{mod} or 
conformal~\cite{conf} invariance may provide a solution to the hierarchy 
problem and to gauge coupling unification. 

Experimental data from LHC will soon decide if the supersymmetric 
particles are at least partly close to the weak scale. If this is not 
the case, then both the MSSM and split supersymmetry are disfavoured. 
Our proposal that superheavy LSPs (SHLSP) are the DM particles
may be then an interesting alternative connecting SUSY to 
the physical world. The prospects to detect DM in the form of stable 
superheavy neutralinos will be discussed in a subsequent work~\cite{next}.

\subsection{Neutralino as LSP}

We assume throughout that the lightest neutralino $\chi\equiv\chi_1$ 
is the lightest of the  supersymmetric particles.
The neutralino mass-matrix $M_\chi$ in the 
$(\tilde{B}, \tilde{W}^0, \tilde{H}^0_1, 
\tilde{H}^0_2)$ basis is given by~\cite{notation}
\begin{equation}\label{E:Mn}
\left(
\begin{array}{cccc} 
 M_1 & 0 & -c_{\beta } m_Z s_W & m_Z s_W s_{\beta } \\
 0 & M_2 & c_W c_{\beta } m_Z & -c_W m_Z s_{\beta } \\
 -c_{\beta } m_Z s_W & c_W c_{\beta } m_Z & 0 & -\mu  \\
 m_Z s_W s_{\beta } & -c_W m_Z s_{\beta } & -\mu  & 0
\end{array}
\right)
\end{equation}
with $s_\beta=\sin\beta$, $c_\beta=\cos\beta$ where $\tan\beta=v_1/v_2$ is
the ratio of the two Higgs vev, $s_W=\sin\theta_W$, $c_W=\cos\theta_W$
with $\theta_W$ as Weinberg angle, and $\mu$ as the Higgs mixing 
parameter. 

The consequences of the limit $M_1,M_2,|\mu| \gg m_Z$ for the neutralino 
has been already extensively discussed for a neutralino as thermal 
relic~\cite{old}. 
Neglecting the terms of order $m_Z$, the four neutralino mass eigenstates 
become a pure bino, wino, and the symmetric and anti-symmetric 
combination of the two higgsinos,
\be
 \{ \tilde{B}, \tilde{W}^0, (\tilde{H}^0_1+\tilde{H}^0_2)/\sqrt{2},
(\tilde{H}^0_1-\tilde{H}^0_2)/\sqrt{2} \}
\ee 
with masses 
\be
 \{ M_1, M_2, -\mu, \mu \} \,.
\ee 
In order to decide which of the two higgsino combinations is the lightest,
one has to include corrections of second order in $m_Z$. Then one obtains 
as mass eigenvalues
\ba
 \Big\{
 M_1-\frac{ s^2_W \left(\mu \sin(2\beta )+M_1\right) m_Z^2}{\mu ^2-M_1^2},
\nonumber\\
 M_2-\frac{c^2_W \left(\mu\sin(2\beta)+M_2\right) m_Z^2}{\mu ^2-M_2^2},
 \nonumber\\
 -\mu +\frac{(\sin (2 \beta )-1) \left(\mu+c^2_W M_1 +s^2_WM_2\right)
   m_Z^2}{2(\mu +M_1)(\mu+M_2)}, 
\nonumber\\
 \mu +\frac{ (\sin (2 \beta )+1)(\mu-c^2_WM_1-s^2_WM_2)m_Z^2}
           {2(\mu -M_1)(\mu-M_2)}
\Big\} \,,
\ea
if the masses are not degenerate.  Depending on the sign of $\mu$, the
symmetric ($\mu<0$) or the anti-symmetric ($\mu>0$) combination of the 
higgsino is the LSP for $|\mu|\ll M_1,M_2$. To be definite, we shall
choose always $\mu>0$ in the following.

If the mass difference between the two lighter of the three mass parameters 
$\mu$, $M_1$ and $M_2$ are small compared
to $m_Z$, a large mixing between the gaugino and the higgsino remains
even in the limit $|\mu|,M\gg m_Z$. This case of a (partially) degeneration 
between the neutralino mass parameters $\mu, M_1$ and $M_2$ was dubbed 
well-tempered neutralino and discussed in Ref.~\cite{ArkaniHamed:2006mb}.

\subsection{Unitarity for superheavy particles}

We illustrate with two explicit examples how superheavy SUSY avoids that
longitudinal gauge bosons become strongly coupled to neutralinos for 
$m_\chi\gg m_Z$. The first case is the annihilation of neutralinos
into fermion pairs in the limit of zero relative velocity $v$.
Then the amplitude consists of sfermion, $Z$ and $A$ exchange, 
${\cal M}={\cal M}_Z+{\cal M}_A+{\cal M}_{sf}$. An inspection of the 
annihilation cross section given e.g.\ in Ref.~\cite{JKG} shows
that the longitudinal component $Z_L$ contributes the term
\be
 (\sigma v)_{Z_L} = \frac{\beta_f}{16\pi}\: \frac{g^4}{c_W^4}\,
                    |O_{11}^{''L}|^2 T_3^2\, \frac{m_f^2}{m_Z^4} 
\ee
to the annihilation cross section into fermions with mass $m_f$, isospin
$T_3$, coupling $O_{11}^{''L}$, and $\beta_f^2=1-m_f^2/m_\chi^2$. 
Assuming that there are no cancellations between ${\cal M}_Z$ and 
${\cal M}_A+{\cal M}_{sf}$ as well as that the factor $O_{11}^{''L}$ is of 
$O(1)$, the annihilation cross section would be independent from $m_\chi$ 
and thus violate perturbative unitarity for $m_\chi\gg m_Z$. 

The resolution of this apparent problem lies partly in the specific form of the
neutralino coupling and mass matrix. The interaction between $Z^0$ and the 
neutralinos are given in the unitary gauge by the Lagrangian~\cite{JKG}
\be
  \mc{L}_{Z \chi^0 \chi^0}=
  \frac{g}{2c_W}Z_\mu[\bar{\chi}_n^0 \gamma^\mu (O_{nm}^{''L}P_L+
   O_{nm}^{''R}P_R)\chi_m^0],
\ee
where $n=m=1$ yields the interaction between $Z^0$ and the LSP. Moreover, 
expressing $O_{ij}^{''L,R}$ by the neutralino mixing matrix elements
gives
\begin{equation}
  O_{nm}^{''L}=-O_{nm}^{''R\ast} =\half (-N_{3n}N_{3m}^\ast + 
  N_{4n}N_{4m}^\ast) \,.
\end{equation} 
Neglecting as always below possible CP violation, the coupling becomes 
simply $\propto (N_{13}^2-N_{14}^2)$ for $m=n=1$. For a completely (anti-) 
symmetric higgsino, this coupling vanishes because of $|N_{13}|=|N_{14}|$,
while a bino and wino LSP have $|N_{13}|=|N_{14}|=0$ in the limit $m_Z\to 0$. 
The leading contribution is thus suppressed by $(m_Z/\Ms)^2$ and given by
\be
O_{11}^{''L}=
\begin{cases}
 \frac{\cos(2\beta) s^2_W m_Z^2}{2(\mu ^2-M_1^2)} \,, 
 &\text{if $M_1\ll M_2,\mu$},
\\ 
 \frac{\cos(2 \beta )c^2_W m_Z^2}{2(\mu^2-M_2^2)}  \,, 
 &\text{if $M_2\ll M_1,\mu$},
\\
 \frac{\cos(2\beta) m_Z^2 \left(c^2_WM_1 +s^2_WM_2-\mu\right)}
      {4\mu\left(\mu -M_1\right) \left(\mu -M_2\right)}  \,, 
 &\text{if $\mu \ll M_1,M_2$}.
\end{cases}
\label{supp}
\ee
Thus this coupling vanishes in the limit $m_\chi/m_Z\to\infty$, because 
the longitudinal components of the gauge bosons couple only to the
deviation from a completely (anti-) symmetric mixing of the higgsino
components. Although the approximations~(\ref{supp}) are valid only for 
non-degenerate masses, this conclusion holds also for degenerate masses, 
because $|N_{13}|\approx|N_{14}|$ remains valid. Taking into
account this suppression factor from the neutralino mixing matrix,
already the single term from $Z_L$ exchange in the annihilation cross section 
is consistent with the unitary bound, $\sigma_{\rm ann}\propto 1/m_\chi^2$.


Another way to understand how the apparently dangerous terms $m_\chi/m_Z$
disappear in physical quantities is to compare the coupling of
the $Z$ and its goldstone boson $G_Z$ in different gauges. For
simplicity, we use for this comparison the unitary and the $R_{\xi}$-gauge 
restricted to  $\xi=1$. Then the $Z$ propagator becomes
purely transversal in the $R_{\xi}$-gauge, and the interactions of 
neutralinos with the goldstone $G_Z$ have to agree with those with the
longitudinal part of the $Z$ boson in unitary gauge.

We consider as example the annihilation of neutralinos into a $Z$ and
the lighter CP even higgs boson in the limit of vanishing relative
velocity $v$. Then the longitudinal part $Z_L$ gives for
$\chi(p)+\chi(p')\to Z(k)+h(k')$ annihilation at rest~\cite{Labonne:2006hk}
\be
 {\cal M}(\chi\chi\rightarrow Zh)_{Z_L} =  
 \frac{-ig^{2}O_{11}^{Z}}{2\cos^{2}\theta_{W}}\frac{m_{\chi}}{m_{Z}}\overline{\chi}\gamma_{5}\chi\frac{q\cdot\eps\left(k\right)}{q^{2}-m_{Z}^{2}} \,,
\label{AmpZxi}
\ee
where $q=p+p'$ denotes the momentum of the virtual $Z_L$ and $\eps$  the 
polarization vector of the real $Z$ boson.
Although the neutral Goldstone boson is part of the $Z$ boson in the 
unitary gauge, its coupling to neutralinos differs and is given by
\be
C_{\chi_n\chi_m}^{G} = \frac{igO_{nm}^{G}}{2\cos\theta_{W}}
\label{G-neutralino-neutralino-coupling}
\ee
with
\be
O_{nm}^{G}=\left(N_{n2}c_{W}-N_{n1}s_{W}\right)\left(c_{\beta}N_{m3}+s_{\beta}N_{m4}\right)+{\scriptstyle \left(n\leftrightarrow m\right)} \,.
\label{OGij}
\ee
For the lightest neutralino annihilation, $n=m=1$, and since the
coupling is imaginary, the Goldstone only couples to the axial part
of the neutralino. The amplitude for the Goldstone exchange diagram
is thus
\be
 {\cal M}(\chi\chi\rightarrow Zh)_{G} =
 i\frac{g^{2}O_{11}^{G}}{2c_{W}^{2}} \overline{\chi}\gamma_{5}\chi\ \frac{q.\varepsilon\left(k\right)}{q^{2}-m_{Z}^{2}} \,.
\label{AmpGxi}
\ee
Comparing (\ref{AmpZxi}) and (\ref{AmpGxi}), gauge independence
requires as relation between the couplings 
\be
 O_{11}^{Z}\frac{m_{\chi}}{m_{Z}} = -\frac{1}{2}O_{11}^{G} \,,
\ee
or expressed in terms of the neutralino mixings and masses,
\ba
\lefteqn{ (N_{14}^{2}-N_{13}^{2})\frac{m_{\chi}}{m_{Z}} =}
\nonumber\\
 &=& - (c_WN_{12}-s_WN_{11})(s_{\beta}N_{14}+c_{\beta}N_{13}) \,.
\label{gauge1}
\ea
The authors of Ref.~\cite{Labonne:2006hk} noted that this relation can be 
derived directly from the definition of the  neutralino mixing matrix $N$,
\be
\left(NM\right)_{nm} =  m_n N_{nm} \,,
\ee
since the identity
\be
 \left(m_n+m_m\right)\left(NPN^{-1}\right)_{nm} =  
 \left(N\left(MP+PM\right)N^{-1}\right)_{nm}
\label{relamatrix}
\ee
holds for any matrix $P$. Choosing for $P$ the isospin operator that
flips the first higgsino sign compared to the second one, 
$P={\rm diag}(0,-\sigma_3)$, reproduces a generalization of Eq.~(\ref{gauge1}).
It is the special structure of the neutralino mass matrix that
makes $PM+MP$ off-diagonal, and thereby leads to the
vanishing of the left-hand side of (\ref{relamatrix}) with $m_{Z}$.


Finally, superheavy particles in a theory with chiral particles may lead
to radiative effects that do not vanish for $m_\chi\to\infty$. 
The authors of Ref.~\cite{Dobado} discussed in a series of works, if the SM 
can be viewed as the low-energy limit of the MSSM in the sense of the 
Appelquist-Carazzone theorem~\cite{Appelquist:1974tg}. They showed that all 
virtual effects of the SUSY particles are either suppressed by inverse powers 
of their mass or can be absorbed in the renormalization
of SM parameters~\footnote{For technical reasons, the authors of 
Ref.~\cite{Dobado} required that the masses of the SUSY particles are not 
degenerate.}.
In conclusion, superheavy SUSY particles do neither
lead to a violation of perturbative unitarity or to non-decoupling effects
in virtual corrections.

\section{Relevant processes and cross sections}

\subsection{Elastic scattering on fermions and the energy relaxation time}

Kinetic equilibrium of neutralinos in the late universe may be reached by 
scattering on light fermions like neutrinos and electrons. In the
rest frame of the neutralino, the Mandelstam variables become
\be
  s=2\omega m_{\chi}+m_{\chi}^2,\quad t=-2\omega^2(1-\cos\theta) \,,
\ee
where $m_{\chi}$ is the mass of the lightest neutralino, $\omega$ is the 
initial energy of the lepton and $\theta$ is the scattering angle. 
We consider here only the case of a broken electroweak symmetry, i.e.\
the case of temperatures $T$ below the weak scale, when the following 
hierarchy holds
\be \label{E:Assumption1} 
 \omega \ll m_Z\ll m_{\chi} \,.
\ee
The assumption  $m_Z\ll m_{\chi}$ leads also to several simplifications 
in the Higgs sector of the MSSM that we shall employ below. Additionally, 
we require that the neutralino mass parameters are not too degenerate,
\be
  |\mu-M_1|,\, |\mu-M_2|,\,  |M_2-M_1|\gg m_Z \,.
\ee
We consider explicitly the case where the lightest neutralino is a bino 
or a higgsino and scatters on a neutrino. The case of a wino is almost 
identical to the one of the bino.

The Feynman amplitudes of the process $\chi+\nu_e\to\chi+\nu_e$
consists of three contributions: 
Sneutrino exchange in the $s$ and $u$ channel, and $t$ channel
exchange of higgses and the $Z$, 
$|\mc{M}|^2 =|\mc{M}_s-\mc{M}_u+\mc{M}_t|^2$.
Since the neutralino is a Majorana particle, the amplitudes $\mc{M}_s$ and
$\mc{M}_u$ can be obtained by interchanging the neutralino in the initial
and final states and thus they should be subtracted.

\subsubsection{The bino as the LSP}
Using the approximations explained above, we obtain as the leading 
contribution to the total spin-averaged squared Feynman amplitude 
in the case of a bino~\cite{comphep}
\be
 |\mc{M}_u|^2  = |\mc{M}_s|^2 =
 \frac{e^4  M_1^2 \omega^2}{2c_W^4 \left(M_{\tilde{\nu}}^2-M_1^2\right)^2} \,,
\ee
\be
 |\mc{M}_t|^2  =
 \frac{e^4 M_1^2\omega^2(3-\cos(\theta))\cos^2(2\beta)}
      {2c_W^4\left(\mu ^2-M_1^2\right)^2} \,,
\ee
\be
2\text{Re}(\mc{M}_s\mc{M}_u^\ast)  = -
 \frac{e^4 M_1^2\omega^2 \sin^2(\theta/2)}
      {c_W^4\left(M_{\tilde{\nu}}^2-M_1^2\right)^2} \,, 
\ee
\ba
\lefteqn{
 2\text{Re}(\mc{M}_s\mc{M}_t^\ast) = - 2\text{Re}(\mc{M}_t\mc{M}_u^\ast) =}
\nn\\ 
 &=&\frac{e^4  \cos(2\beta) M_1^2\omega^2 (3-\cos(\theta ))}
       {2c_W^4\left(M_1^2-M_{\tilde{\nu}}^2\right) \left(M_1^2-\mu ^2\right)} \,.
\ea
Here, we used neutrinos as scattering target and denoted by $M_{\tilde\nu}$
the sneutrino mass.  

The energy relaxation time can be calculated as (see e.g.\ Ref.~\cite{bde})
\begin{align}
\label{E:enreltimedef}
  \frac{1}{\tau_\text{rel}}=
  \frac{N_{\rm eff}}{2E_k m_{\chi}}\int_0^\infty \!\!\!d\omega \int d\Omega\:
  n_0(\omega)(\delta p)^2 \left(\frac{d\sigma_\text{el}}{d\Omega}\right)_{f_L\chi} ,
\end{align}
where $E_k=(3/2)T$ is the mean kinetic energy of the neutralinos, 
$\delta p$ the neutralino momentum obtained in one scattering,
\begin{align}
  (\delta p)^2=2\omega^2(1-\cos(\theta))
\end{align}
and the number density of relativistic fermions with one polarization and 
energy $\omega$ is
 \begin{align} 
  n_0=\frac{1}{2\pi^2}\frac{\omega^2}{e^{\omega/T}+1}\approx
  \frac{1}{2\pi^2}\:\omega^2 \, e^{-\omega/T} \,.
 \end{align}
Finally, the factor $N_{\rm eff}$ counts the number of relevant relativistic
degrees of freedom, weighted with the relative size of their cross-section
compared to a neutrino. Combining the different contributions and 
performing the integrals gives
\begin{equation}
\label{E:enreltimeBino}
 \tau_{\text{rel}} =
 \frac{\pi^3 c_W^4 M_1 \left(M_{\tilde{\nu}}^2-M_1^2\right)^2 
       \left(\mu ^2-M_1^2\right)^2}
      {25N_{\rm eff} e^4 T^6 \left[\cos(2\beta) M_{\tilde{\nu}}^2 +
       \mu ^2-2 c^2_\beta M_1^2\right]^2} \,.
\end{equation}

\subsubsection{The higgsino as the LSP}
In the case of a  higgsino as the LSP the contributions to the total squared
Feynman amplitude $|\mc{M}|^2$ are given by
\be
 |\mc{M}_s|^2  = 
 \frac{2e^4 \mu^2\omega^2 m_Z^4 \left(M_1 c_W^2+M_2 s_W^2-\mu \right)^4 
        \left(c_\beta+s_\beta\right)^4}
      {s_{2 W}^4(\mu -M_1)^4 
        \left(\mu -M_2\right)^4 \left(M_{\tilde{\nu}}^2-\mu ^2\right)^2} \,,
\ee
\be
 |\mc{M}_u|^2 =
 \frac{2e^4 \mu^2 \omega^2 m_Z^4 \left(M_1 c_W^2+M_2 s_W^2-\mu \right)^4 
        \left(c_\beta+s_\beta\right)^4}
      {\left(\mu -M_1\right)^4 \left(\mu -M_2\right)^4 
       \left(M_{\tilde{\nu}}^2-\mu ^2\right)^2} \,,
\ee
\be
 |\mc{M}_t|^2 = 
 \frac{2e^4 \omega^2 c_{2 \beta }^2 \left(3-\cos\theta\right) 
       \left(M_1 c_W^2+M_2 s_W^2-\mu\right)^2}
      {s_{2 W}^4\left(\mu -M_1\right)^2 \left(\mu -M_2\right)^2} \,,
\ee
\ba
 \lefteqn{2\text{Re}(\mc{M}_s\mc{M}_t^\ast) = 
 2 e^4 \left(c_{2 \beta }+s_{4 \beta }/2\right) \left(3-\cos\theta\right)}
\nonumber\\
& \times &
\frac{\mu  m_Z^2 \omega ^2 \left(-M_1 c_W^2-M_2 s_W^2+\mu
   \right)^3 }{s_{2 W}^4\left(\mu -M_1\right)^3 \left(\mu
   -M_2\right)^3 \left(\mu ^2-M_{\tilde{\nu }}^2\right)} \,,
\ea
\ba
 \lefteqn{2\text{Re}(\mc{M}_s\mc{M}_u^\ast) = 
 - 4 e^4\left(c_{\beta }+s_{\beta }\right)^4 \sin ^2(\theta/2)}
\nonumber\\
&\times &
\frac{ \mu ^2 \omega ^2  m_Z^4 \left(M_1 c_W^2+M_2
   s_W^2-\mu \right)^4 }{s_{2 W}^4\left(\mu -M_1\right)^4 \left(\mu
   -M_2\right)^4 \left(M_{\tilde{\nu }}^2-\mu ^2\right)^2}  
\ea
\ba
 \lefteqn{2\text{Re}(\mc{M}_t\mc{M}_u^\ast) = -2 e^4 \left(c_{2 \beta }+s_{4\beta} /2\right) \left(3-\cos\theta \right)}
\nonumber\\
& \times &
   \frac{ \mu  \omega ^2  m_Z^2 \left(-M_1 c_W^2-M_2 s_W^2+\mu
   \right)^3 }{s_{2 W}^4\left(\mu -M_1\right)^3 \left(\mu -M_2\right)^3 \left(\mu ^2-M_{\tilde{\nu }}^2\right)} \,.
\ea
Analogously to the case of the bino, the energy relaxation time for a Higgsino 
as lightest neutralinos follows as
\be
\tau_{\text{rel}} =\frac{\pi ^3 \mu ^3 \left(\mu -M_1\right)^2 \left(\mu -M_2\right)^2 s_{2 W}^4}{100 N_{\rm eff} e^4 T^6 c_{2
   \beta }^2 \left(M_1 c_W^2+M_2 s_W^2-\mu \right)^2}.
\ee

\subsection{Elastic neutralino-neutralino scattering}
We use in this subsection again the assumptions~\eqref{E:Assumption1}, 
but denote now with $\omega$ the kinetic energy of the colliding
neutralinos in their center of mass frame. Then the Mandelstam variables 
are
\begin{equation}
  s=4(M_{\chi}^2+\omega^2),\qquad t=-2\omega^2(1-\cos\theta) \,.
\end{equation}

\subsubsection{The Bino as the LSP}
Neutralino-neutralino scattering  can occur through $s$ and $t$ channel 
exchange of the $Z$ and the three neutral Higgs bosons. We shall see
that it is sufficient to consider only the squared matrix elements. 
They are given in the unitary gauge by
\begin{align}
|\mc{M}_{\text{$Z$-exch}}|^2 &= \frac{9 e^4 \cos ^4(2 \beta ) m_Z^4 M_1^4 \tan ^4\left(\theta _W\right)}{2 \left(\mu ^2-M_1^2\right)^4} \,, 
\\
|\mc{M}_{\text{$Z$-ann}}|^2 &= \frac{e^4 \cos ^4(2 \beta ) m_Z^4 M_1^4 
\tan ^4\left(\theta _W\right)}{2 \left(\mu ^2-M_1^2\right)^4} \,,
\\  
|\mc{M}_{\text{$h$-ann}}|^2 &=\frac{e^4 \omega ^4 m_Z^4 \left(\mu  
\sin (2 \beta )+M_1\right)^4 \tan ^4\left(\theta
    _W\right)}{2M_1^4 \left(\mu ^2-M_1^2\right)^4}\\
|\mc{M}_{\text{$h$-exch}}|^2 &=\frac{8 e^4 m_Z^4 M_1^4 \left(\mu  
\sin (2 \beta )+M_1\right)^4 \tan ^4\left(\theta _W\right)}{M_h^4
   \left(\mu ^2-M_1^2\right)^4}\label{E:h-exch} \,,
\\
|\mc{M}_{\text{$H$-exch}}|^2 &= \frac{8 e^4 \mu ^4 \cos ^4(2 \beta ) m_Z^4 M_1^4 \tan ^4\left(\theta _W\right)}{M_H^4 \left(\mu
   ^2-M_1^2\right)^4} \,
\\
|\mc{M}_{\text{$H$-ann}}|^2 &=\frac{8 e^4 \mu ^4 \omega ^4 
\cos ^4(2 \beta ) m_Z^4 \tan ^4\left(\theta _W\right)}{\left(M_H^2-4
   M_1^2\right)^2 \left(\mu ^2-M_1^2\right)^4} \,
\\
|\mc{M}_{\text{$A^0$-ann}}|^2 &= \frac{8 e^4 m_Z^4 M_1^4 \left(\mu +\sin (2 \beta ) M_1\right)^4 \tan ^4\left(\theta _W\right)}{\left(M_{A^0}^2-4 M_1^2\right)^2
   \left(\mu ^2-M_1^2\right)^4} 
\end{align}
and
\ba
\lefteqn{|\mc{M}_{\text{$A^0$-exch}}|^2 = e^4 \tan^4\left(\theta_W\right)
}
\\
&&
\frac{\omega ^4 (\cos (2 \theta )+7) m_Z^4 \left(\mu +\sin (2 \beta ) M_1\right)^4}{M_{A^0}^4 
 \left(\mu ^2-M_1^2\right)^4}
\,.
\nonumber
\label{E:H3-exch}
\ea
We note first that we can neglect the squared amplitudes proportional to 
$\omega^4$. Because of the hierarchy in the Higgs masses,
\begin{equation}\label{E:order1}
 \mc{O}(M_h)=\mc{O}(m_Z) \ll \mc{O}(M_H) =\mc{O}(M_{A^0}) \,.
\end{equation}
the $h$-exchange channel~\eqref{E:h-exch} that is of order 
$\mc{O}(M_{SUSY}^0)$ compared to other channels of $\mc{O}(m_Z^4/M_{SUSY}^4)$ 
dominates the self-scattering of superheavy neutralinos. With 
$|\mc{M}_{\chi^0\chi^0 \to \chi^0\chi^0}|^2 = |\mc{M}_{\text{$h$-exch}}|^2$,  
the total cross section of neutralino-neutralino scattering follows
as
\be
\sigma=\frac{e^4 m_Z^4 M_1^2 \left(\mu  \sin (2 \beta )+M_1\right)^4 \tan ^4\left(\theta _W\right)}{16 M_h^4 \pi  \left(\mu ^2-M_1^2\right)^4}.
\ee

\subsubsection{The Higgsino as the LSP}
Analogously to the bino case, the leading contribution to higgsino-higgsino
scattering is given by the exchange of the light, SM-like Higgs $h$.
With
\ba
\lefteqn{
  |\mc{M}_{\chi^0\chi^0 \to \chi^0\chi^0}|^2 = |\mc{M}_{\text{$h$-exch}}|^2 =
  \frac{e^4\left(c_{\beta }+s_{\beta }\right)^8}{2 
   c_W^4 s_W^4}}
\nonumber
\\ &&
\frac{ \mu ^4 m_Z^4 \left(M_1 \cos ^2\left(\theta _W\right)-\mu +\sin ^2\left(\theta _W\right) M_2\right)^4 }{ M_h^4 \left(\mu -M_1\right)^4 \left(\mu -M_2\right)^4}\,
\ea
the total cross section of neutralino-neutralino scattering follows
as
\be
 \sigma=\frac{e^4  \left(c_{\beta }+s_{\beta }\right)^8\mu ^2 m_Z^4 \left(M_1 c_W^2+M_2 s_W^2-\mu \right)^4}{256 \pi  c_W^4s_W^4M_h^4 \left(\mu -M_1\right)^4 \left(\mu -M_2\right)^4
   }.
\ee

\subsection{Annihilations}
The annihilations of neutralinos have been studied in great detail. 
Annihilations of superheavy neutralinos are, in the bino case, dominated by the
channels $Z H,\: h A,\:A H, \: W^\pm H^\mp$, since all other channels
are suppressed by powers of $m_Z/M_{\rm SUSY}$.
In the higgsino case all bosonic channels contribute at leading order
except annihilation into $Z^0+A^0$ and $h+H$. Annihilation into fermions are
always suppressed.

The fermionic annihilation 
channels do not give leading order contributions in any case.

\subsubsection{The bino as the LSP}
Assume that $\{M_{\chi_1},M_{\chi_2},M_{\chi_3},M_{\chi_4}\}$ correspond
to $\{M_1,M_2,\mu,-\mu\}$. Then the squared matrix elements of these 
dominant channels are given by
\ba\label{E:annIntoZH}
\lefteqn{|\mc{M}_{\chi^0\chi^0\to Z^0 H}|^2 =e^4 \left(M_H^2-4 M_1^2\right)^2\times} 
\\
&&
\frac{\left(-4 \mu  M_1^3+A+B+\mu ^2 M_A^2 s_{2 \beta }\right)^2}{8 c_W^4 \left(\mu
   ^2-M_1^2\right)^2 \left(M_A^2-4 M_1^2\right)^2 \left(-2 \mu ^2-2 M_1^2+M_H^2\right)^2}\nn
\,,
\ea
with
\ba
A &=&\left(M_A^2-2 M_H^2\right) s_{2 \beta } M_1^2\\
B &=& 2
   \mu  \left(2 \mu ^2+M_A^2-M_H^2\right) M_1 \,,
\ea
\ba\label{E:annIntoZHA}
\lefteqn{|\mc{M}_{\chi^0\chi^0\to A^0 H}|^2= e^4 \left(M_A^2-M_H^2\right)^2
 /(8 c_W^4 t_{2 \beta }^2) \: \times }
\\
&&\frac{ \left(-8 s_{2 \beta } M_1^4-8 \mu  M_1^3+C+D+E\right)^2}{ \left(\mu ^2-M_1^2\right)^2 \left(M_A^2-4
   M_1^2\right)^2 \left(-2 \mu ^2-2 M_1^2+M_A^2+M_H^2\right)^2 }
 \,,
\nonumber
\ea
with 
\ba
C &=&2 \left(M_A^2+M_H^2\right)
   s_{2 \beta } M_1^2\\
D &=& 4 \mu  \left(-2 \mu ^2+M_A^2+M_H^2\right) M_1 \\
E &=& c_{\beta } M_A^2 \left(-4 \mu
   ^2+M_A^2+M_H^2\right) s_{\beta }
\,,
\ea
\begin{align}
 |\mc{M}_{\chi^0\chi^0\to h A^0}|^2 =
 \frac{e^4 \left(\sin (2 \beta ) M_A^2+4 \mu  M_1\right)^2}
      {8 c_W^4\left(M_A^2-2 \mu ^2-2 M_1^2\right)^2}  \,.
\end{align}
and
\ba\label{E:annIntoWpmHpm}
\lefteqn{|\mc{M}_{\chi^0\chi^0\to W^\pm H^\mp}|^2 = 
 e^4 \left(M_{H^\pm}^2-4 M_1^2\right)^2/8 c_W^4 \: \times}
\\ 
&&\frac{\left(-4 \mu  M_1^3+F+G+\mu ^2 M_A^2 s_{2 \beta
   }\right)^2}{ \left(M_{H^\pm}^2-2 \mu ^2-2 M_1^2\right)^2 \left(\mu ^2-M_1^2\right)^2
   \left(M_A^2-4 M_1^2\right)^2}
\nn \,.
\ea
where
\ba
F &=&\left(M_A^2-2 M_{H^\pm}^2\right) s_{2
   \beta } M_1^2\\
G &=& 2 \mu  \left(-M_{H^\pm}^2+2 \mu ^2+M_A^2\right) M_1 \,.
\ea
The annihilation cross section of the various channels follows then as
\be
 \sigma v = \frac{\beta_f}{32\pi m_\chi^2} \sum_i |M_i|^2 \,.
\ee

\subsubsection{The higgsino as the LSP}
Assume that $\{M_{\chi_1},M_{\chi_2},M_{\chi_3},M_{\chi_4}\}$ correspond
  to $\{\mu,-\mu,M_1,M_2\}$.
Then the squared matrix-elements of these dominant channels are given
by
\ba\label{E:annIntoZZhiggsino}
|\mc{M}_{\chi^0\chi^0\to Z^0 Z^0}|^2 =\frac{2 e^4}{\sin{(2 \theta_W)}^4}
\,,
\ea
\ba\label{E:annIntoWWhiggsino}
|\mc{M}_{\chi^0\chi^0\to W^+ W^-}|^2 =\frac{e^4}{4 \sin(\theta_W)^4}
\,,
\ea
\ba\label{E:annIntoZhhiggsino}
\lefteqn{|\mc{M}_{\chi^0\chi^0\to Z^0 h}|^2 =\frac{2 e^4 \mu ^2 c_{2 \beta }^2 }{\left(\mu ^2+M_1^2\right)^2 \left(\mu ^2+M_2^2\right)^2 s_{2 W}^4}\: \times}
\\ 
&&
\left(\mu ^2 M_2 c_W^2+M_1^2 M_2 c_W^2+M_1 \left(\mu ^2+M_2^2\right)
   s_W^2\right)^2 \nn
\,,
\ea
\ba\label{E:annIntoAhhiggsino}
\lefteqn{|\mc{M}_{\chi^0\chi^0\to A^0 h}|^2 =\frac{e^4  \left(c_{\beta }+s_{\beta
   }\right)^4}{2 \left(-2 \mu ^2-2 M_1^2+M_A^2\right)^2 A^2 s_{2 W}^4}\: \times}\nn
\\ 
&&
 \left(4 \mu  M_1 s_W^2 A-2 c_W^2 M_1^2 B+4 \mu 
   c_W^2 M_2 C+D\right)^2 \,,
\ea
where
\ba
A &=& \left(-2 \mu ^2-2 M_2^2+M_A^2\right)   \\
B &=& \left(M_A^2+4 \mu  M_2\right) \\
C &=& \left(M_A^2-2 \mu ^2\right) \\
D &=& M_A^4-2 \mu ^2 M_A^2-2 M_2^2 s_W^2 M_A^2
\,,
\ea
The annihilation into $W^\pm, H^\mp$ are dominated by the chargino
exchange and the $A^0$ annihilation channels. In order to obtain 
relatively compact expressions,  we set $M_{A^0}=M_{H^\pm}$ and 
assume that $\mu  \cos (\beta )+\sin (\beta ) M_2\ne 0$. As
a result we obtain
\ba\label{E:annIntoWHhiggsino}
|\mc{M}_{\chi^0\chi^0\to W^\pm H^\mp}|^2 =
  \frac{e^4 \left(s_{2\beta} M_{H^\pm}^2+4\mu M_2\right)^2}
  {8\left(-2\mu ^2-2 M_2^2+M_{H^\pm}^2\right)^2 s_W^4}.
\ea
To simplify the squared matrix elements both for annihilations 
into $Z^0+H$ and $A^0+H$, we set $M_{A^0}=M_H$. The results are
\ba\label{E:annIntoZHhiggsino}
\lefteqn{|\mc{M}_{\chi^0\chi^0\to Z^0 H}|^2 =
  \frac{e^4  \left(c_{\beta }-s_{\beta }\right){}^4}{H \left(-2 \mu ^2-2 M_2^2+M_H^2\right){}^2 } \: \times \nn}
  \\
  &&
  \left(M_H^4-2 \mu ^2 M_H^2+E-F-G\right){}^2 \,, 
\ea
with
\ba
E &=&  4 \mu  M_1 \left(2 \mu ^2+2 M_2^2-M_H^2\right) s_W^2 \\
F &=& 2 c_W^2 M_1^2 \left(M_H^2-4 \mu  M_2\right) \\
G &=&  2 M_2 \left(2 \mu  \left(M_H^2-2 \mu ^2\right)
   c_W^2+M_2 M_H^2 s_W^2\right) \\
H &=& 2 s_{2 W}^4 \left(-2 \mu ^2-2 M_1^2+M_H^2\right){}^2
\,,
\ea
and
\ba\label{E:annIntoAHhiggsino}
\lefteqn{|\mc{M}_{\chi^0\chi^0\to A^0 H}|^2 =
 \frac{2 e^4 \mu ^2 c_{2 \beta }^2 }{\left(\mu ^2+M_1^2-M_H^2\right){}^2
   I}  \: \times \nn}
   \\
   && 
  \left(M_1^2 M_2 c_W^2+M_2 \left(\mu ^2-M_H^2\right) c_W^2+J \right){}^2
 \ea  
with
\ba
I &=&\left(\mu ^2+M_2^2-M_H^2\right){}^2 s_{2 W}^4 \\
J &=& M_1 \left(\mu ^2+M_2^2-M_H^2\right) s_W^2 
\,.
\ea
The  annihilation channels  $Z^0+A^0$ and $h+H$ do not give an 
$\mathcal{O}(1)$ contribution.

\section{Cosmological production}
All the mechanisms for the production of SHDM particles described in 
section~\ref{SHDM} apply also to SHLSPs. In particular, 
the abundance of neutralinos due to gravitational production is 
given by Eq.~(\ref{abund}).  
In most works on SHDM particles  the dependence on $T_R$ and 
the two-fold degeneracy in Eq.~(\ref{abund}) was fixed by 
choosing first for $T_R$ the highest value allowed by the gravitino problem, 
$T_R=10^9$\,GeV. Then the larger of the two possible masses was selected,
$m_\chi\sim 3\times 10^{13}$\,GeV, so that secondaries of SHDM decays 
could explain the cosmic rays of the highest energies, $E\gsim 10^{20}$\,eV.

Both constraints can be relaxed in our scenario: Superheavy gravitinos 
decay before big-bang nucleosynthesis and we do not insist that SHDM 
decays explain UHECRs with energies, $E\gsim 10^{20}$\,eV. As a result, 
the only constraint additionally to Eq.~(\ref{abund}) is the requirement
that neutralinos do not thermalize, $T_R\ll m_\chi/30$.
Thus the choice $T_R=10^{10}$\,GeV and $m_\chi=3\times 10^{11}$\,GeV  
gives the smallest possible neutralino mass.
Finally we note that decays of heavier SUSY particles with mass $\tilde m$, 
that are more effectively produced in the regime $\tilde m\ll H_I$ than 
neutralinos, can change the relation~(\ref{abund}) but not
the smallest possible neutralino mass.
\vspace*{0.1cm}

\section{Summary}
We have suggested the lightest supersymmetric particle as a well-suited 
candidate for superheavy dark matter. The requirement that SHDM is produced 
with the correct abundance by gravitational interactions at the end of 
inflation leads to a lower mass limit of $3\times 10^{11}\,{\rm GeV}$  
for the masses of all SUSY particles. Since weak interactions remain 
perturbative despite the large mass hierarchy, $m_\chi\gg m_Z$, and the mass 
scales are approximately fixed, we could reliably calculate the relevant
processes for the special case of a superheavy neutralino.

Our proposal can be falsified in the near future by the discovery
of low-scale/split SUSY at the LHC. If this is not the case, then 
SHLSPs as DM particles may be the unique opportunity to connect SUSY 
to the physical world. The observable consequences of meta-stable 
SHDM were discussed already in detail in 
Refs.~\cite{Aloisio:2006yi,Aloisio:2007bh}, while the prospects to detect 
DM in the form of stable superheavy neutralinos despite of their small 
number density and annihilation cross section will be discussed in a 
subsequent work~\cite{next}.

\acknowledgments
%
We would like to thank T.~Plehn for helpful discussions and 
especially A.~Pukhov for advice on the use of CalcHEP.


\end{document}